\documentstyle[aps,prl,multicol]{revtex}

\begin{document}
\input{epsf}
\draft
\preprint{}
\title{Signatures of spin pairing in a quantum dot in the Coulomb 
blockade regime}
\author{ S. L\"uscher$^{1}$,  T. Heinzel$^{1}$, K. Ensslin$^{1}$,W. 
Wegscheider,$^{2,3}$ and M. Bichler$^{3}$}
\address{$^{1}$Solid State Physics Laboratory, ETH Z\"{u}rich, 8093
Z\"{u}rich,  Switzerland\\
$^{2}$Institut f\"ur Angewandte und Experimentelle Physik, Universit\"at 
Regensburg, 93040 Regensburg, Germany\\
$^{3}$Walter Schottky Institut, TU M\"unchen, 85748 Garching, 
Germany}
\date{\today}
\maketitle
\begin{abstract}
Coulomb blockade resonances are measured in a GaAs quantum dot in which both shape 
deformations and interactions are small. The parametric evolution of 
the
Coulomb blockade peaks shows a 
pronounced 
pair correlation in both position and amplitude, which is interpreted as spin pairing. 
As a consequence, the nearest-neighbor distribution of peak spacings 
can be well approximated by a smeared bimodal Wigner surmise, 
provided that
interactions which go beyond the 
constant interaction model are taken into account . 
\end{abstract}

\begin{multicols}{2}
\narrowtext
Recently, the Coulomb blockade (CB) of electronic transport through quantum 
dots, defined in two-dimensional electron gases in semiconductor 
heterostructures, has been of considerable interest \cite{Kouwenhoven97}. 
One reason is that such dots are model systems to investigate the interplay 
between chaos and 
electron-electron (e-e) interactions. Here, a key feature is the 
distribution of nearest-neighbor Coulomb blockade peak spacings 
(NNS), which random matrix 
theory \cite{Mehta91} (RMT) predicts to follow a bimodal Wigner surmise 
$P(s)$ for a non-interacting 
quantum dot of chaotic shape, i.e.
\begin{equation}
P(s)=\frac{1}{2}\left[\delta(s)+P^{\beta}(s)\right]
\end{equation}
 $P^{\beta}(s)$ is the 
 Wigner surmise for the corresponding Gaussian ensemble, i.e. $\beta 
 =1$ for systems with time inversion symmetry (Gaussian orthogonal 
 ensemble - GOE), and $\beta =2 $ when time inversion symmetry is broken 
 (Gaussian unitary ensemble - GUE). The peak spacing s is measured in units of the 
 average 
  spin-degenerate energy level spacing $\Delta= 
2\pi \hbar^{2}/ (m^{*}A)$, where $m^{*}$ denotes the effective mass, 
and A the dot area.  
The $\delta$-function in $P(s)$ takes 
the spin degeneracy into account. RMT further 
 predicts the standard deviation for $P(s)$ to be $\sigma = 0.62$ for  $\beta 
 =1$, and $\sigma =0.58$ for $\beta =2$, respectively \cite{Beenakker97}.\\
 The comparison to experimental data is made by applying the 
 constant-interaction model \cite{Beenakker91,Kastner92}, which allows to separate the constant 
 single-electron charging energy from the fluctuating energies of the 
 levels
 inside the dot. In disagreement with the predictions of RMT, the experimentally obtained NNS 
 distributions are usually best 
 described by a {\it single Gaussian} with enhanced values of $\sigma$ 
 \cite{Sivan96,Simmel97,Patel98a,Simmel99}.  The data thus look as if 
 spins are absent, although in Ref. \cite{Patel98b}, a spin pair has 
 been observed. This 
 apparent absence of spins and the different shape of $P(s)$ have triggered tremendous
 recent theoretical work.
 One possible explanation
 are additional e-e interactions inside the dot
 \cite{Sivan96,Berkovits97,Blanter97,Cohen99,Walker99,Ahn99,Alhassid99}, 
 which lead to 
 ``scrambling'' of the energy spectrum \cite{Stewart97,Patel98b} 
 and can be characterized by the interaction parameter 
 $r_{s}$, defined as the ratio 
 between the Coulomb interaction of two electrons at their average 
 spacial separation, and the Fermi energy\cite{Berkovits97,Walker99,Ahn99,Alhassid99}. 
 It is 
 theoretically expected that the NNS distribution becomes Gaussian due 
 to e-e interactions, and that $\sigma$ increases for $ r_{s} \geq 2$. 
 \cite{Walker99,Ahn99}. However, all experiments 
 so far have been 
 carried out in a regime 
 where an increase in $\sigma$ is not expected, 
 i.e. in samples with $0.93\leq r_{s}\leq 
 1.35$ \cite{Sivan96,Simmel97,Patel98a},
 with the exception of Ref. \cite{Simmel99}, where $r_{s}= 2.1$.\\
 Gate-voltage induced shape deformations of the dot can modify the NNS distribution as well.
 The deformation can be described by a parameter $x$, which 
 corresponds to the distance between avoided crossings induced by the 
 deformation, measured in units of the CB peak spacing. For $x 
 \approx$ 1, the NNS distribution of partly uncorrelated energy 
 spectra is measured, resulting again in a Gaussian shape 
 with enhanced $\sigma$\cite{Hackenbroich97,Vallejos98}. Whether 
 shape deformations or interactions dominate the shape of the NNS 
 distribution is not clear, although there is experimental evidence 
 that $x <1$ and interactions are more 
 important\cite{Simmel99,Patel98b}.\\
 Here, we report measurements on a 
 quantum dot in which shape deformations as well as $r_{s}$ 
 are reduced.  We observe a pronounced pairwise correlation of both 
 position and amplitude of the Coulomb blockade resonances, which is 
 sometimes interrupted by kinks in the parametric evolution, among 
 other features. We interpret the pairing as a spin signature: the energies of 
 two states belonging to the same spatial wave function with opposite 
 spin differ by an average interaction energy $\bar\xi$, which fluctuates 
 with a standard deviation of $\sigma_{\xi}$, both of which are of the 
 order of $\Delta$. We conclude that in 
 previous experiments, spin pairing was difficult to observe because 
 it was frequently destroyed by avoided level crossings. Furthermore, we 
 suggest that the measured NNS distribution can be fitted to a modified bimodal Wigner 
 surmise, with $\bar\xi$ and $\sigma_{\xi}$ as fit parameters. \\ 
 The sample is a shallow Ga[Al]As heterostructure with 
 a two-dimensional electron gas (2DEG) $34$ nm below the surface. The 
 quantum dot is defined by local oxidation with an atomic force 
 microscope \cite{Held98} (inset in Fig.1(a)). The 
 lithographic dot area is 280 nm x 280 nm. The dot can be 
 tuned by voltages applied to a homogeneous top gate and to the planar 
 gates I and II. In order to reduce $r_{s}$ as much as possible, we 
 chose a heterostructure with a high electron density, further increased by a top gate 
 voltage of $+100$ mV to $n_{e}=5.9 \cdot 10^{15}$ m$^{-2}$. This 
 results in $r_{s}=0.72$, which is smaller than in all previous 
 experiments. Additional screening is provided by the top gate 
 \cite{TH1}. The sample was mounted in the mixing chamber of a
 $^{3}$He/$^{4}$He-dilution refrigerator with a base temperature of
 90\,mK. The mobility of the cooled 2DEG was $93$ m$^{2}$/Vs. A DC bias 
 voltage of  10 $\mu V$ was applied across the dot, and the current is measured with 
 a resolution of 500 fA. From capacitance 
 measurements\cite{Kastner92}, we find the electronic dot area A=190 nm x 190 nm (the depletion length in 
 such devices can be smaller than in structures defined by top 
 gates\cite{Held98}) . The single-electron 
 charging energy is $E_{c}$ =1.25 meV and the spin-edgenerate level 
 spacing $\Delta$ = 200 $\mu eV$.
  \begin{figure}
      \centerline{\epsfxsize=8.2cm \epsfbox{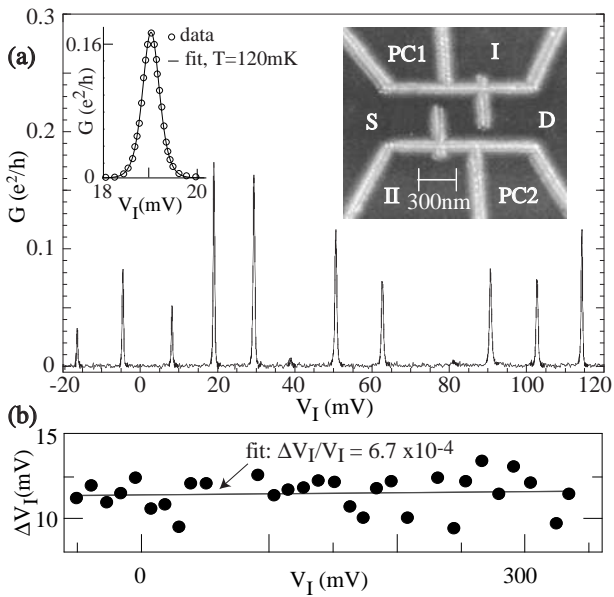}}
      \caption[Fig. 1] {(a) Right inset: AFM picture (taken before evaporation 
     of the top gate) of the oxide 
     lines (bright) that define the dot, coupled to source (S) and 
     drain (D) via tunnel barriers, which can be adjusted with the planar gates 
     PC1 and PC2.  
     Gates I and II are used to tune the dot. Main figure: 
     Conductance  G as a 
     function of  $V_{I}$, showing Coulomb blockade resonances. Left inset:  fit (line) to 
     one measured CB peak (open circles), see text.
     (b) Linear fit (line) of the peak spacing $\Delta V_{I}$as a function of 
     $V_{I}$ (dots). The average peak spacing is almost constant, 
     indicating small shape deformations.}
     \label{SL1}
 \end{figure}
The measurements have been carried out in the weak coupling regime,
$\hbar\Gamma\ll k_{B}T \ll \Delta$. Here, $\Gamma$ 
denotes the coupling of the dot to source and drain. The 
conductance G was measured as a function of the voltage $V_{I}$ applied 
to the planar gate I (see inset in Fig. 1(a)). Magnetic fields B applied perpendicular to the 
sample surface and $V_{II}$ were used as parameters.
 The observed CB oscillations (Fig. 1(a)) are fitted to a thermally 
 broadened line shape, i.e., $G(V_{I})=G_{max}cosh^{-2}(\eta V_{I}/2k_{B}T)$ \cite{Beenakker91}, 
 yielding an electron temperature of T=120mK, 
 as well as the positions and amplitudes of the peaks. Here, $\eta 
 =0.11 eV/V$ is the lever arm.  
 Fig. 1 (b) shows 
 the peak spacing $\Delta V_{I}$ as a function of $V_{I}$. Compared to 
 conventional dots defined by top gates, 
 we find a much smaller variation of the average peak spacing as $V_{I}$ is 
 tuned, although the fluctuation of individual spacings is 15\% of  $E_{c}$. 
 A linear fit gives a slope of $\Delta V_{I}/V_{I}=6.7\cdot 10^{-4}$.
 Hence, the capacitance between the dot and gate I varies only 
 by 3\% over the whole scan range, as compared to, for example, a factor of 3 in Ref. 
 \cite{Simmel97}. This indicates that tuning gate I or II 
 predominantly changes the energy of the conduction band bottom, 
 while the dot is only slightly deformed. By applying the method 
 of Ref. \cite{Hackenbroich97} to a hard-wall confinement, we 
 estimate $x \approx$ 0.15 for our dot as a lower limit. 
 \\
  In Fig. 2 (a), five consecutive CB 
 peaks are shown as a function of B. A pronounced 
 pairwise correlation of both 
 amplitude and peak position is observed (peak b correlates with peak 
 c, and peak d with peak e, respectively). 
  Observation of a pairing has been reported previously and 
  interpreted as spin pairing \cite{Patel98b}, but not been further investigated.\\
  We interpret this parametric pair correlation in terms of a model recently developed 
 by Baranger et al. \cite{Baranger99}. The constant interaction 
 model is used to subtract $E_{c}$ from the peak spacings. The remaining individual 
 energy separations equal $\Delta/2$ on average and reflect the 
 fluctuating level separations inside 
 the quantum dot, which consist of two parts. We assume that two 
 paired peaks belong to the same spatial wave function, labelled by i, of opposite 
 spin, and are split by an interaction energy $\xi_{i}$, 
 while the energy of consecutive states with 
 different orbital wave functions differs by $\Delta _{i}- \xi_{i}$. 
 Since the separations between the two levels of equal spin of spin 
 pair i and (i+1), $\Delta_{i}$, and possibly also $\xi_{i}$, vary as a function 
 of B, levels may cross and the ground state of the dot can be either a 
 singlet or a triplet state.  At the singlet-triplet transitions, 
 kinks in the parametric peak evolution occur and the pair 
 correlation is interrupted\cite{Baranger99}. We can identify such 
 kinks in our data, among other features. Fig. 2 (b)
 shows the amplitudes of peaks
c, d and e. The correlation between peaks d and e is very strong around B=0.  For 
0.4T$<$B$<$0.61T, this correlation is interrupted, while the 
amplitudes of peaks c and 
e are correlated instead.  In this regime, 
correlated kinks in the evolution of peaks c and d are observed 
(Fig.2 (c)).  In Fig. 2 (d), a possible 
corresponding scenario for the parametric dependence of energy levels 
is sketched: (left) two avoided crossings occur between level pair i and 
level pair i+1 . This leads to the position of peaks c, d, and e 
as sketched in Fig. 2 (d), right, corresponding to the difference in 
energy upon changing the electron number in the dot. Consequently, 
positions and amplitudes  of peaks c and e should be correlated in 
0.4T$<$B$<$0.61T, as observed. Note that this correlation is interrupted 
around B=0.5T, possibly due to the influence of another energy level.\\
Also, $\xi_{de}$ is not constant over the full range of B. 
While $\xi_{de}\approx 0.05\Delta$ for B$<$0.22T, the positions of 
peaks d and e are not detectable in 0.22T$<$B$<$0.32T, since their 
amplitudes vanish. As the peaks reappear, $\xi_{de}$ has jumped to 
$\xi_{de}^{*}\approx 0.25 \Delta.$ We speculate that possibly a level crossing has 
occurred in the regime where the amplitudes are suppressed, and hence 
for B$<$0.22T, a different level pair is at the Fermi energy than for 
B$>$0.32T. In addition, we note that although $\xi$ fluctuates as B is 
varied, a systematic change of $\xi$ with B is not observed, which
indicates that Zeemann splitting plays a minor role. 
From the data of Fig. 2, we estimate the average 
 interaction energy to 
 $\overline{\xi} \approx 0.5 \Delta$ by averaging over all peaks and 
 magnetic   
 fields. Baranger et al. have estimated 
 $\overline{\xi}\approx 0.6 \Delta$ for $r_{s}=1$. Hence, our findings can be 
 considered as being in agreement with existing theory, 
 while we are not aware of a theoretical prediction for 
 $\sigma_{\xi}$. From the 
 above phenomenology, we conclude that for dots with stronger 
 shape deformations, and hence more level crossings, or in dots with 
 larger $r_{s}$ (and thus larger $\overline{\xi}$), the spin pairing is frequently 
 interrupted and difficult to detect. Also, the Kondo effect 
 \cite{Goldhaber98} can occur 
 in neighboring peaks \cite{Maurer99} when spin pairing is interrupted. \\ 
 We proceed by discussing the effect of spin pairing on the NNS 
 distributions. 
 In Fig.  3, the measured histograms of the normalized NNS 
distributions for GOE (a) and GUE (b) are shown.  
The ensemble statistics have been obtained by measuring $G(V_{I})$, 
 and  either by changing   the 
 magnetic flux  by one flux quantum $\phi_{o}=\frac h e$ through the dot (GUE), or by 
 stepping $V_{II}$ in units of one CB period (GOE). Each 
 individual $V_{I}$-sweep contains 15 CB resonances in the low 
 coupling regime. 
 The total number of peak spacings used is 120 for GOE, and 210 for 
 GUE, respectively. The individual level spacings $s$ in units of $\Delta$ 
 are obtained by using the fit 
of Fig. 1b; its expectation value is $\overline{s}$=0.5. 
Both histograms are asymmetric and show no 
evident bimodal structure. By including the effect of spin pairing 
into the statistics, however, we can interpret them as modified bimodal 
distributions:\\
(i) The $\delta$-function in the non-interacting 
NNS distribution $P(s)$ with the expectation value of 
$\overline{s_{\delta}}$=0  (eq. (1)) is 
shifted to $\overline{s_{\delta}}=\overline{\xi^{*}}$ and, as a reasonable 
assumption \cite{Baranger00}, broadened according to a Gaussian 
distribution with the standard deviation $\sigma_{\xi^{*}}$. Here, $\xi^{*}$ denotes the interaction energy in units of $\Delta$.\\
(ii) Since one level of a spin pair i is shifted upwards in energy by 
$\xi_{i}¥$,  the separation between the upper level of spin pair i 
and the lower level of pair (i+1) is given by $\Delta_{i}$-$\xi_{i}$. 
Consequently, $P^{\beta}(s)$ in eq. (1)
is shifted to $\overline{s_{P^{\beta}}}$=1-$\overline{\xi^{*}}$ and convoluted with the 
Gaussian distribution function of $\xi^{*}$.\\
Combining these two components, the modified 
NNS distribution reads\\ 
\begin{figure}
      \centerline{\epsfxsize=8.0cm \epsfbox{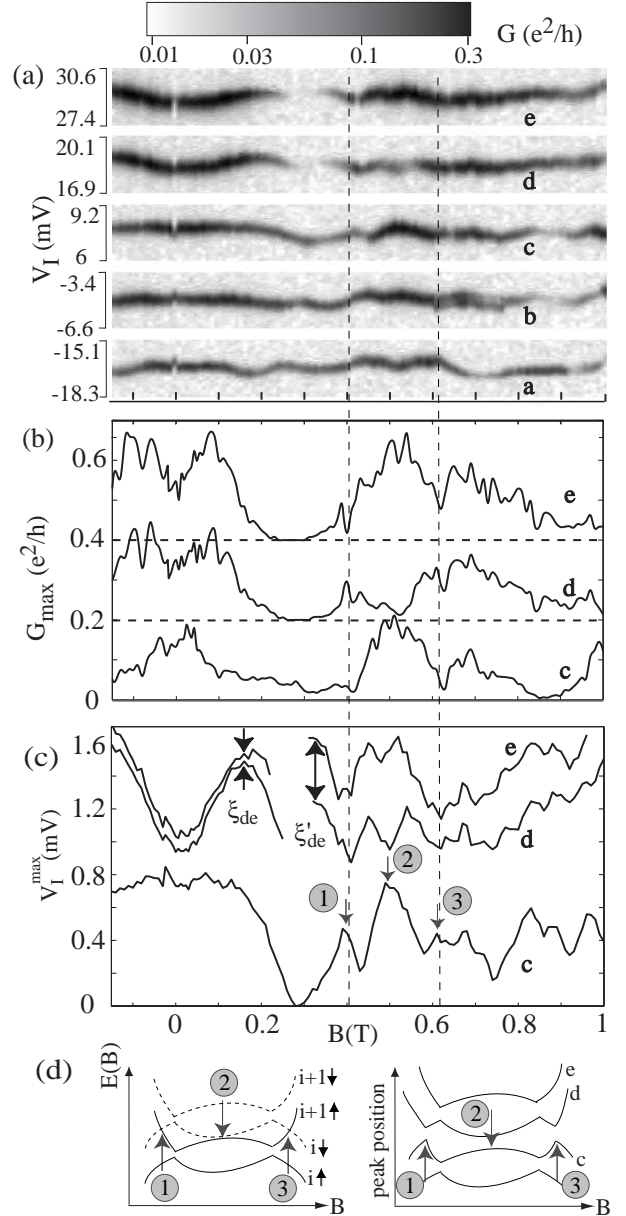}}
     \caption[Fig. 2] {(a) Logarithmic grayscale plot of paramatric 
     variations in a magnetic field B for 5 
     consecutive CB peaks.  
     A pair correlation in peak position amplitude is observed, which is interrupted in 
     certain ranges of B, for example in the region between the dashed 
     lines.
     (b) Parametric amplitudes for peaks c,d, and e, offset by 0.2 $e^{2}/h$ 
     each. The correlation between peak d and e is lost in 
     0.4T$<$B$<$0.6T, and  e correlates with c instead. 
     (c) The corresponding position of the peak maxima. The 
     traces are offset for clarity. At magnetic fields labelled by 
     1 and 3, kinks in the peak position occur, while the 
     separation between peak d and e jumps across the region of 
     suppressed amplitude from $\xi_{de}$ to $\xi'_{de}$. (d) Scheme of a 
     possible double anticrossing between spin-paired level i and 
     i+1 (left, the black arrows indicate the spin), which could lead to the observed structure in the 
     correlation for peaks c, d and e (right).}
     \label{bild2}
 \end{figure}
 $P_{int}^{\beta}(\overline{\xi^{*}}, \sigma_{\xi^{*}})=$
\begin {equation}
\frac {1} {\sqrt{2\pi} \sigma_{\xi^{*}}}\left\{ exp\left[ -\frac {(s-\overline{\xi^{*}})^2}
{2\sigma_{\xi^{*}}^{2}}\right] + exp\left[ -\frac {s^2} 
{2\sigma_{\xi^{*}}^2}\right] \times 
P^{\beta}(s+\xi^{*})\right\}
\end {equation}
Here, the ``$\times$'' denotes the convolution.
Since $\Delta$ is determined by the dot size and the material 
parameters, we can fit $P_{int}^{\beta}(\overline{\xi^{*}}, \sigma_{\xi^{*}})$ 
 \begin{figure}
      \centerline{\epsfxsize=8.0cm \epsfbox{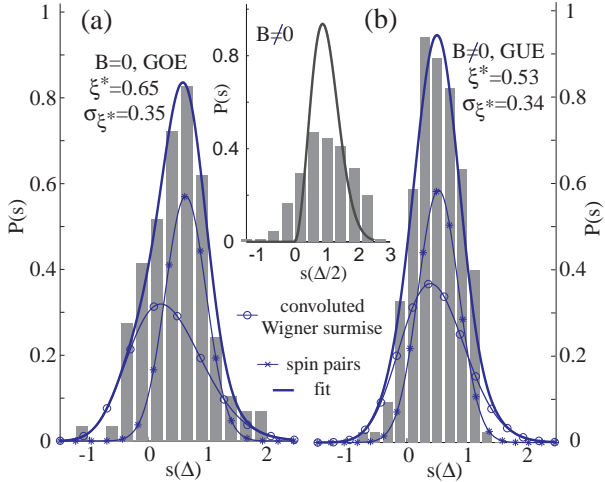}}
     \caption[Fig. 3] {Measured NNS distributions (gray bars) for B=0 (a) 
     and B $\neq$ 0 (b). The bold solid curves are the fits to 
     $P_{int}^{\beta}(\overline{\xi^{*}},\sigma_{\xi^{*}})$, with the fit 
     results as indicated in the figure (see text).  Also drawn are 
     the two components of $P_{int}^{\beta}$, i.e. the Gaussian 
     distribution of separations between spin pairs, and its 
     convolution with the corresponding Wigner surmises. The inset 
     compares the GUE data to $P^{2}(s)$ (eq. 1), using 
     the spin-resolved level spacing $\Delta/2$ as the average peak 
     separation. 
       }
\label{bild3}
 \end{figure}
to the measured NNS distribution 
with the two fit parameters $\overline{\xi^{*}}$ and $\sigma_{\xi^{*}}$ 
(Fig.3). We 
obtain $\overline{\xi^{*}}=0.65$ and $\sigma_{\xi^{*}} 
=0.35$ for GOE, as well as $\overline{\xi^{*}} =0.53$ and $\sigma_{\xi^{*}}  
=0.34$ for GUE. 
Hence, we find that $\overline{\xi}$ is higher for GOE than for GUE, 
which is in agreement the theoretical prediction \cite {Baranger99}. 
The fluctuation of $\xi$ is found to be independent of the Gaussian 
ensemble, and does not vary continously with B\\
In these fits, we have assumed that two electrons are always 
successively filled in 
one spatial wave function, i.e. we have neglected situations in 
which $\xi_{i}>\Delta_{i}$. Inclusion of avoided crossings
would require more clearly pronounced kinks 
than those in our data (sometimes the pair correlation is lost 
while a kink is not clearly visible). More 
experiments as well as theoretical work is necessary to investigate 
this dependence, also with respect to fluctuations in $\xi$ with B.\\
  Finally, we consider how our data can be modelled when complete 
  absence of spin pairing is assumed. In this case, the mean level 
  spacing would be $\Delta/2$. Comparing a correspondingly normalized 
  Wigner surmise to our data gives an extremely poor result (inset 
  in Fig.3): the measured NNS distribution appears too wide by a 
  factor of $\approx$ 2.\\
In summary, we have observed spin pairing effects in a - compared to 
dots investigated in earlier experiments - rigid quantum  dot 
with reduced electron-electron interactions. 
We have observed spin pairing which persists as a magnetic field is varied, but 
 is interrupted by kinks as well as other structures in the parametric evolution of the Colomb 
 blockade peaks. We have extracted the average interaction 
 energy $\overline{\xi}$ between states of identical spatial wave functions but 
 opposite spin. Furthermore, we explain 
 the measured distributions of nearest neighbor spacings as being composed of the two branches of a modified, bimodal 
 Wigner-Dyson distribution, which takes $\overline{\xi}$ and its 
 fluctuation into 
 account.\\ 
 It is a pleasure to thank H. U. Baranger, E. Mucciolo, K. Richter,  and F. Simmel for stimulating 
 conversations and discussions. Financial support from the Schweizerischer Nationalfonds is 
 gratefully acknowledged.

\end{multicols}

\begin{references}
\bibitem{TH1}A simple image charge calculation reveals that the top 
gate reduces $r_{s}$  only by $\approx 2\%$. Nevertheless, it improves 
screening  of impurity potentials and  the confining potential becomes steeper.
\end{references}
\end{document}